\documentclass [10pt]{article}

\begin{document}
\begin{flushright}
DIAS-STP-04-05
\end{flushright}
\begin{center}
{\bf Exact Solution of Noncommutative  $U(1)$ Gauge Theory in  $4-$Dimensions }\\
\bigskip
{\bf Badis Ydri}
\bigskip

{\it School of Theoretical Physics, \\
Dublin Institute for Advanced Studies, Dublin, Ireland.}\\

\end{center}
\begin{abstract}

Noncommutative $U(1)$ gauge theory on the Moyal-Weyl space ${\bf
R}^2{\times}{\bf R}^2_{\theta}$ is regularized  by approximating
the noncommutative spatial slice ${\bf R}^2_{\theta}$ by a fuzzy
sphere of matrix size $L$ and radius $R$ . Classically we observe
that the field theory on the fuzzy space ${\bf R}^2{\times}{\bf
S}^2_L$ reduces to the field theory on the Moyal-Weyl plane ${\bf
R}^2{\times}{\bf R}^2_{\theta}$ in the flattening continuum
planar limits $R,L{\longrightarrow}{\infty}$ where the ratio
${\theta}^2=R^2/|L|^{2q}$ is kept fixed with  $q>\frac{3}{2}$ . The effective
noncommutativity parameter is found to be given by
${\theta}_{eff}^2{\sim}2{\theta}^2(\frac{L}{2})^{2q-1}$ and thus
it corresponds to  a strongly noncommuting space. In the quantum
theory it turns out that this prescription is also equivalent to a
dimensional reduction of the model where the noncommutative
$U(1)$ gauge theory  in $4$ dimensions is shown to be equivalent
in the large $L$ limit to an ordinary $O(M)$ non-linear sigma
model in $2$ dimensions where $M{\sim}3L^2$ . The Moyal-Weyl model
defined this way is also seen to be an ordinary renormalizable
theory which can be solved exactly using the method of steepest
descents . More precisely we find for a fixed  renormalization
scale $\mu$ and a fixed renormalized coupling constant $g_r^2$ an
$O(M)-$symmetric mass , for the different components of the sigma
field , which is non-zero for all values of $g_r^2$ and hence the
$O(M)$ symmetry is never broken in this solution . We obtain also
an exact representation of the beta function of the theory which
agrees with the known one-loop perturbative result .
\end{abstract}

\section{Introduction }

We propose in this article to reconsider the problem of quantum
$U(1)$ gauge theory in $4-$dimensions where spacetime is
noncommutative . In particular we will consider the simple case
where only two spatial directions are noncommutative and thus
avoiding potential problems with unitarity and causality. Towards
the end of regularizing this model we replace the noncommutative
Moyal plane with a fuzzy sphere , i.e with a $(L+1){\times}(L+1)$
matrix model where $a=\frac{1}{L+1}$ is essentially a
lattice-spacing-like parameter . The fuzzy sphere ${\bf S}^2_L$
has two cut-offs , a UV cut-off $L$ ( the matrix size ) and an IR
cut-off $R$ ( the radius ) which both  preserve Lorentz, gauge
and chiral symmetries, and which allows us to view the
noncommutative Moyal plane as a sequence of matrix models
$Mat_{L+1}(R),. ..,Mat_{L^{'}+1}(R^{'}),...,Mat_{L^{"}+1}(R^{"})$
with the two parameters $L$ and $R$ ever increasing (
$L...{\leq}L^{'}...{\leq}L^{"}$, $R...{\leq}R^{'}...{\leq}R^{"}$
) while ( for example ) the ratio $R/L={\theta}/2$ is kept fixed.
In this way one can immediately see that Lorentz symmetry is only
lost at the strict limit in the sense that the original $SO(3)$
symmetry is reduced to an $SO(2)$ symmetry while the
noncommutativity parameter ${\theta}^2$ in this prescription is
equal to the volume of spacetime per point ( in here this is given
by the area of the sphere divided by the number of points , i.e
${\pi}{\theta}^2=\frac{4{\pi}R^2}{L^2}$).

In this section we will first recall few results from
noncommutative perturbative gauge theory which will be useful to
us in what will follow in this article \cite{nekrasov,szabo} . The
basic noncommutative gauge theory actions of interest to us in
this article are matrix models of the form \cite{nekrasov}
\begin{eqnarray}
S_{\theta}=\frac{{\theta}^d}{4g^2}Tr\hat{F}_{ij}^2=\frac{{\theta}^d}{4g^2}Tr\sum_{i,j}\bigg(i[\hat{D}_i,\hat{D}_j]-\frac{1}{{\theta}^2}(B^{-1})_{ij}\bigg)^2.\label{action}
\end{eqnarray}
$i$ , $j=1,...,d$ , $B^{-1}$ is assumed in here to be an invertible tensor ( which in
$2$ dimensions is
$(B^{-1})_{ij}=({\epsilon}^{-1})_{ij}=-{\epsilon}_{ij}$ ) , and
${\theta}$ has dimension of length so that the operators
$\hat{D}_i$'s have dimension of $({\rm length})^{-1}$. The
coupling constant $g$ is of dimension $(\rm
mass)^{2-\frac{d}{2}}$ . The trace is taken over some infinite
dimensional Hilbert space ${\bf H}$ and hence
$Tr[\hat{D}_i,\hat{D}_j]$ is ${\neq}0$ in general . In general
$Tr$ is equal to the tarce over coherent states ( corresponding
to spacetime )  times the trace over the gauge group if any ( in
here this is simply $U(1)$ ) . The sector of this matrix theory
which corresponds to a noncommutative gauge field on ${\bf
R}^d_{\theta}$ is defined by the configurations \cite{nekrasov}
\begin{eqnarray}
\hat{D}_i=-\frac{1}{{\theta}^2}(B^{-1})_{ij}\hat{x}_j+\hat{A}_i,~\hat{A}_i^{+}=\hat{A}_i,\label{cond1}
\end{eqnarray}
where the components $\hat{x}_i$'s can be identified with those of
a background noncommutative gauge field whereas $\hat{A}_i$'s are
identified with the components of the dynamical $U(1)$
noncommutative gauge field  . $\hat{x}_i$'s can also be
interpreted as the coordinates on the noncommutative space ${\bf
R}^d_{\theta}$ satisfying the usual commutation relation
\begin{eqnarray}
[\hat{x}_i,\hat{x}_j]=i{\theta}^2B_{ij}.\label{cond2}
\end{eqnarray}
Derivations on this ${\bf R}^d_{\theta}$ will be taken for
simplicity to be defined by
\begin{eqnarray}
\hat{\partial}_i=-\frac{i}{{\theta}^2}(B^{-1})_{ij}\hat{x}_j=-\hat{\partial}_i^{+}~,~[\hat{{\partial}_i},\hat{x}_j]={\delta}_{ij}~,~[\hat{{\partial}_i},\hat{{\partial}_j}]=\frac{i}{{\theta}^2}(B^{-1})_{ij}.\label{cond3}
\end{eqnarray}
$U(1)$ gauge transformations which leave the action (\ref{action})
invariant are implemented by unitary matrices
$U=exp(i{\Lambda})~,~UU^{+}=U^{+}U=1~,~{\Lambda}^{+}={\Lambda}$
which act on the Hilbert space ${\bf H}$ as follows . The
covariant derivative
$\hat{D}_i=-i\hat{\partial}_i+\hat{A}_i^aT_a$ and curvature
$\hat{F}_{ij}=i[\hat{D}_i,\hat{D}_j]-\frac{1}{{\theta}^2}(B^{-1})_{ij}=[\hat{\partial}_i,\hat{A}_j]-[\hat{\partial}_j,\hat{A}_i]+i[\hat{A}_i,\hat{A}_j]$
transform as $\hat{D}_i{\longrightarrow}U\hat{D}_iU^{+}$ ( i.e
$\hat{A}_i{\longrightarrow}U\hat{A}_iU^{+}-iU[\hat{\partial}_i,U^{+}]$
) , $\hat{F}_{ij}{\longrightarrow}U\hat{F}_{ij}U^{+}$. By virtue
of (\ref{cond1}) , (\ref{cond2}) and (\ref{cond3}) it is not
difficult to show that the matrix action (\ref{action}) is
precisely the usual noncommutative gauge action on ${\bf
R}^d_{\theta}$ with a star product defined by the parameter
${\theta}^2B^{ij}$ , i.e
\begin{eqnarray}
S_{\theta}=\frac{1}{4g^2}\int d^dx
{F}_{ij}^2~;~{F}_{ij}={\partial}_i{A}_j-{\partial}_j{A}_i+i\{{A}_i,{A}_j\}_{*}.\label{**}
\end{eqnarray}
Quantization of the matrix models (\ref{action}) consists usually
in quantizing the models (\ref{**}) . This generally makes good
sense at one-loop but not necessarily at higher loops which we
still do not know how to study systematically . Let us
concentrate in the rest of this introduction on the $U(1)$ model
in $d=4$. The one-loop effective action can be easily obtained (
for example )  in the Feynamn-'t Hooft background field gauge where $A_i=A_i^{(0)}+A_{i}^{(1)}$ 
one finds the result \cite{martin}
\begin{eqnarray}
{\Gamma}_{\theta}=S_{\theta}[A^{(0)}]-\frac{1}{2}Tr_dTRLog\bigg(({\cal
D}^{(0)})^2{\delta}_{ij}+2i{\cal F}_{ij}^{(0)}\bigg)+TRLog({\cal
D}^{(0)})^2,
\end{eqnarray}
where the operators $({\cal D}^{(0)})^2={\cal D}_i^{(0)}{\cal
D}_i^{(0)}$ , ${\cal D}_i^{(0)}$ and ${\cal F}_{ij}^{(0)}$ are
defined through a star-commutator and hence even in the $U(1)$
case ( which is of most interest in here anyway )  the action of
these operators is not trivial, viz for example ${\cal
D}_i^{(0)}(A_j^{(1)}){\equiv}[D_i^{(0)},A_j^{(1)}]_{*}=-i{\partial}_iA_j^{(1)}+[A_i^{(0)},A_j^{(1)}]_{*}$
, etc . $Tr_d$ is the trace associated with the spacetime index
$i$ and $TR$ corresponds to the trace of the different operators
on the Hilbert space ${\bf H}$ . As an illustrative example we
compute now explicitly the quadratic effective action . This will
also contain all quantum corrections to the vacuum polarization
tensor . After a long calculation \cite{martin} one obtains
\begin{eqnarray}
{\Gamma}_{\theta}^{(2)}=-\frac{1}{2}\int
\frac{d^d\vec{p}}{(2{\pi})^d} \bigg[(p^2{\delta}_{ij
}-p_{i}p_{j})\bigg(\frac{1}{g^2}+{\Pi}^{P}(p)\bigg)+{\Pi}_{ij}^{NP}(p)\bigg]A_{i}^{(0)}(p)A_{j}^{(0)}(-p).
\end{eqnarray}
Explicitly we find in particular that the planar function is UV
divergent as in the commutative theory and thus requires a
renormalization. Indeed by integrating over arbitrarily high
momenta in the internal loops we see that the planar amplitude
diverges so at any arbitrary scale ${\mu}$ one finds in $d=4 +2
\epsilon $ the closed expression \cite{martin}
\begin{eqnarray}
{\Pi}^{P}(p)&=&-\frac{11}{24{\pi}^2}(\frac{1}{\epsilon}+\gamma
+ln\frac{p^2}{{\mu}^2}
)+\frac{1}{24{\pi}^2}+\frac{1}{8{\pi}^2}\int_0^1dx(1-2x)^2~lnx(1-x)-\frac{1}{2{\pi}^2}\int_0^1dx~lnx(1-x).
\end{eqnarray}
Obviously in the limit $\epsilon{\longrightarrow}0$ this planar
amplitude diverges , i.e their singular high energy behaviour is
logarithmically divergent . These divergent contributions needs
therefore a renormalization . Towards this end it is enough as it
turns out to add the following counter term to the bare action
\begin{eqnarray}
{\delta}S_{\theta}=-\frac{1}{4}\big(\frac{11}{24{\pi}^2\epsilon}\big)\int
d^dx F_{ij}^{(0)2}.
\end{eqnarray}
The claim of \cite{martin,wul} is that this counter term will
also substract the UV divergences in the $3-$ and $4-$point
functions of the theory at one-loop and hence the theory is
renormalizable at this order. The vacuum polarization tensor at
one-loop is therefore given by
\begin{eqnarray}
{\Pi}_{ij}^{1-loop}&=&(p^2{\delta}_{ij}-p_{i}p_{j})\frac{1}{{g}^2(\mu)}+{\Pi}_{ij}^{NP}(p).
\end{eqnarray}
where

\begin{eqnarray}
\frac{1}{g^2(\mu)}
&=&\frac{1}{g^2}-\frac{11}{24{\pi}^2}ln\frac{p^2}{{\mu}^2}-\frac{11}{24{\pi}^2}{\gamma}+\frac{1}{24{\pi}^2}+\frac{1}{8{\pi}^2}\int
dx\big[(1-2x)^2-4\big]lnx(1-x).
\end{eqnarray}
A starightforward calculation gives then the beta function
\begin{eqnarray}
{\beta}(g(\mu))={\mu}\frac{d{g(\mu)}}{d{\mu}}=-\frac{11}{24{\pi}^2}g^3(\mu).\label{beta}
\end{eqnarray}
We remark on the other hand that the non-planar function
${\Pi}_{ij}^{NP}(p)$ is finite in the UV because of the presence
of              a      regulating exponential of the form
$exp(-\frac{\tilde{p}^2}{4t})$ in loop integrals where
$\tilde{p}_i={\theta}^2B_{ij}p_j$ . However it is obvious that
this noncommutativity-induced exponential regularizes  the
behaviour at high momenta ( which corresponds to the values
$t{\longrightarrow}0$ ) only when the external momentum
$\tilde{p}$ is ${\neq}0$. Indeed in the limit of small
noncommutativity or small momenta we have the infrared singular
behaviour
\begin{eqnarray}
{\Pi}_{ij}^{NP}(p)&=&\frac{11}{24{\pi}^2}(p^2{\delta}_{ij}-p_{i}p_{j})ln~p^2\tilde{p}^2+\frac{2}{{\pi}^2}\frac{\tilde{p}_{i}\tilde{p}_{j}}{(\tilde{p})^2}.\label{pro}
\end{eqnarray}
This also means that the renormalized vacuum polarization tensor
diverges in the infrared limit $\tilde{p}{\longrightarrow}0$
which is the definition of the UV-IR mixing of  this model.

In this article we will give a nonperturbative  exact
representation  of the beta function (\ref{beta}) in the regime of
strong noncommutativity using the method of large $N$ matrix
models .We will show in particular that the noncommutative $U(1)$
gauge theory  is equivalent to an ordinary large non-linear sigma
model and that the result (\ref{beta}) is actually valid to all
orders in $g$ . We postpone however the discussion of the UV-IR
mixing problem (\ref{pro}) and its solution to a future
communication \cite{denjoeydri1}.

\section{The Fuzzy Sphere As a Regulator of The Moyal-Weyl Plane} As a warm up we will only consider in this
section the case of two dimensions and then go through the
$4-$dimensional case in more detail in  next sections. The action
(\ref{action}) reads in two dimensions as follows
\begin{eqnarray}
S_{\theta}=\frac{{\theta}^2}{4g^2}Tr\hat{F}_{ij}^2=\frac{{\theta}^2}{4g^2}Tr\sum_{i,j}\bigg(i[\hat{D}_i,\hat{D}_j]-\frac{1}{{\theta}^2}({\epsilon}^{-1})_{ij}\bigg)^2.\label{action1}
\end{eqnarray}
The major obstacles in systematically quantizing the above action
(\ref{action1}) are $1)$ the infinite dimensionality of the Fock
space on which the trace $Tr$ is defined , $2)$ the presence of a
dimension-full parameter ${\theta}$ in the theory and $3)$ the
absence of Lorentz invariance because of the existence of a
background magnetic field $B_{ij}$ [ This last point is of course
not relevant in the special case of $2$ dimensions ] .

The above three problems are immediately solved by redefining the
above action as certain limit of finite dimensional matrix models
. Indeed in the case $d=2$ ( which is of most interest to us in
this first section ) we replace (\ref{action1})  by the
$(L+1)-$dimensional matrix model
\begin{eqnarray}
S_{L,R}=\frac{R^2}{4g_f^2}\frac{1}{L+1}Tr_fF_{ab}^2=\frac{R^2}{4g_f^2}\frac{1}{L+1}Tr_f
\sum_{a,b}\bigg(i[D_a,D_b]+\sum_{c}\frac{1}{R}{\epsilon}_{abc}D_c\bigg)^2,\label{action2}
\end{eqnarray}
with the constraint \cite{nair,denjoydri}
\begin{eqnarray}
D_aD_a=\frac{|L|^2}{R^2}~,~|L|^2=\frac{L}{2}(\frac{L}{2}+1).\label{constraint1}
\end{eqnarray}
Now $a,b,c$ take the values $1,2,3$ which means that the above
regularization is effectively embedded in $3$ dimensions and
hence the need for the extra constraint  . The tensor
${\epsilon}_{abc}$ is the $\epsilon$ symbol in $3$ dimensions .
The trace $Tr_f$ is now defined on a finite dimensional Hilbert
space , this trace is dimensionless and the dimension of $({\rm
length})^2$ which is carried by ${\theta}^2$ in
 (\ref{action1}) is now carried by $R^2$ . The equations of motion
derived from the action (\ref{action2}) are given by
\begin{eqnarray}
{\delta}S_{L,R}=\frac{R^2}{g_f^2}\frac{1}{L+1}Tr_f{\delta}D_c\bigg(-i[F_{cb},D_b]+\frac{1}{2R}{\epsilon}_{abc}F_{ab}\bigg){\equiv}0
{\Leftrightarrow}-i[F_{cb},D_b]+\frac{i}{2R}{\epsilon}_{abc}F_{ab}=0.
\end{eqnarray}
An important class of solutions to these equations of motion are
given by the solutions to the zero-curvature condition $F_{ab}=0$
together with the constraint (\ref{constraint1}) . These are the
famous so-called fuzzy spheres and they are essentially defined
by the covariant derivatives $D_a=\frac{L_a}{R}$ for which
$F_{ab}=0$ and $D_aD_a=\frac{|L|^2}{R^2}$ where of course $L_a$'s
are the generators of the $(L+1)-$dimensional irreducible
representation of $SU(2)$. It is also well established
\cite{alex,denjoydri,steinacker,jun} that these solutions are classically stable for
finite $L$ only  because of the constraints (\ref{constraint1})
which we chose in here to impose rigidly [ we could have instead
chosen to implement these constraints in a variety of different
ways  as discussed in \cite{denjoydri,steinacker} ] . We replace
therefore the configurations (\ref{cond1}) by
$(L+1){\times}(L+1)$ matrices given by
\begin{eqnarray}
D_a=\frac{1}{R}L_a+{A}_a.
\end{eqnarray}
The noncommutative coordinates $\hat{x}_i$'s are replaced by the
noncommutative matrix coordinates $x_a=\frac{R}{|L|}L_a$'s
satisfying
\begin{eqnarray}
[{x}_a,{x}_b]=i\frac{R}{|L|}{\epsilon}_{abc}x_c~,~\sum_{a}x_a^{2}=R^2,.\label{cond4}
\end{eqnarray}
Hence we have effectively regularized the noncommutative plane
(\ref{cond2}) with a fuzzy sphere of radius $R$. This can also be
seen as follows . We introduce the $(L+1){\times}(L+1)$ gauge
field and write $F_{ab}=F_{ab}^{(0)}+i[A_a,A_b]$,
$F_{ab}^{(0)}=\frac{i}{R}\big([L_a,A_b]-[L_b,A_a]-i{\epsilon}_{abc}A_c\big)$.
The Yang-Mills action (\ref{action2}) in the large $L$ limit
becomes
\begin{eqnarray}
S_{L,R}{\longrightarrow}\frac{R^2}{4g_f^2}\int
\frac{d{\Omega}}{4{\pi}}(F_{ab}^{(0)})^2.
\end{eqnarray}
As one can immediately see this is indeed the $U(1)$ action on
ordinary ${\bf S}^2$ with radius $R$  and coupling constant $
g_f^2$ .

However in the matrix model (\ref{action2}) we want to think of
$R$ and $L$ as being infrared and ultraviolet cut-offs
respectively of the theory (\ref{action}) with the crucial
property that for all finite values of these cut-offs gauge
invariance and Lorentz invariance are preserved .

The limit in which the finite dimensional matrix model
(\ref{action2}) reduces to the infinite dimensional  matrix model
(\ref{action}) is a continuum  double scaling limit of large $R$
and large $L$ taken as follows
\begin{eqnarray}
R,L~{\longrightarrow}{\infty}~;~{\rm
keeping}~\frac{R^2}{|L|^{2q}}=~{\rm
fixed}{\equiv}{\theta}^2\label{limit}
\end{eqnarray}
with $q$ a real number and where we have also to constrain the
fuzzy coordinate $x_3$ ( for example via the application of an
appropriate projector or by any other means ) to be given by
\begin{eqnarray}
x_3=R.{\bf 1}.\label{lorentz}
\end{eqnarray}
This means that we are effectively restricting the theory around
the north pole of the fuzzy sphere where in the limit of large
$R$ and large $L$ one can reliably set $x_3=R.{\bf 1}$  . The
noncommutative coordinates can then be identified as
$\hat{x}_i=\frac{1}{|L|^{q-\frac{1}{2}}}({\epsilon}^{-1})_{ij}x_j$
or $x_i=|L|^{q-\frac{1}{2}}{\epsilon}_{ij}\hat{x}_j$ with the
correct commutation relations (\ref{cond2}) , i.e
$[\hat{x}_i,\hat{x}_j]=i{\theta}^2{\epsilon}_{ij}$ . Furthermore
by dividing $R^2$ across the identity $\sum_{a}x_a^{2}=R^2$ one
finds the trivial result $1=1$ which means in particular that the
two coordinates $\hat{x}_1$ , $\hat{x}_2$ are now not
constrainted in any way . The traces $Tr_f$ and $Tr$ are on the
other hand  identified in the planar limit through the simple
equation $Tr=Tr_f$ \cite{stern}. Furthermore in this large planar
limit (\ref{limit}) the constraint (\ref{constraint1}) takes the
form
\begin{eqnarray}
\frac{1}{R}\{x_3,A_3\}+\frac{1}{R}\sum_{i=1}^2\{x_i,A_i\}+\theta
|L|^{q-1}\sum_{a=1}^3A_a^{2}=0{\Leftrightarrow}\frac{2}{\theta}\hat{A}_3+\frac{1}{|L|}\hat{A}_3^2+\frac{1}{{\theta}^2|L|}\{\hat{x}_i,\hat{A}_i\}+\frac{1}{|L|}\hat{A}_i^2=0.
\end{eqnarray}
In other words $\hat{A_3}=0$ in this limit and thus one finds
$\hat{D_3}=\frac{1}{{\theta}}$ where $\hat{A}_3=|L|^{q-1}A_3$ and
$\hat{D}_3=|L|^{q-1}D_3$. Clearly we are also using the fact that
we have in this limit $\hat{D}_i=|L|^{q-\frac{1}{2}}D_i$ and
$\hat{A}_i=|L|^{q-\frac{1}{2}}A_i$ where of course
$\hat{D}_i=-\frac{1}{{\theta}^2}({\epsilon}^{-1})_{ij}\hat{x}_j+\hat{A}_i$
.  As a consequence we conclude that
$F_{ij}=\frac{1}{|L|^{2q-1}}\hat{F}_{ij}$ where
$\hat{F}_{ij}=i[\hat{D}_i,\hat{D}_j]-\frac{1}{{\theta}^2}({\epsilon}^{-1})_{ij}$
and
$F_{i3}=-\frac{1}{{\theta}|L|^{2q-\frac{1}{2}}}{\epsilon}_{ij}\hat{D}_{j}$
.  The sum ${\sum}_{ab}$ reduces effectively to $\sum_{ij}$, i.e
the difference
$S_{L,R}-\frac{1}{L+1}\frac{1}{|L|^{2q-2}}S_{\theta}$ being equal
to $\frac{1}{2g_f^2|L|^{2q-1}(L+1)}Tr\sum_{i}(\hat{D}_i)^2$
consists only of vanishing terms of order $\frac{1}{L^{2q}}$ and
hence the action (\ref{action2}) is seen to tend to
(\ref{action1}) with an effective classical coupling given by
\begin{eqnarray}
g_{eff}^2=g_f^2{\xi}^2~,~{\xi}^2=|L|^{2q-2}(L+1).
\end{eqnarray}
However and since we have
\begin{eqnarray}
S_{\theta}=\frac{{\theta}^2}{4g_{eff}^2}Tr\bigg([\hat{D}_i,\hat{D}_j]-\frac{1}{{\theta}^2}({\epsilon}^{-1})_{ij}\bigg)^2=\frac{{\theta}_{eff}^2}{4g_f^2}Tr\bigg([\frac{1}{\xi}\hat{D}_i,\frac{1}{\xi}\hat{D}_j]-\frac{1}{{\theta}_{eff}^2}({\epsilon}^{-1})_{ij}\bigg)^2,
\end{eqnarray}
we can view (\ref{action2}) as describing ( in the limit ) a gauge
theory on a noncommutative plane with an effective
noncommutativity parameter given by
\begin{eqnarray}
{\theta}_{eff}^2={\theta}^2{\xi}^2.\label{effe}
\end{eqnarray}
From here we can conclude that for $q>\frac{1}{2}$ ,
${\xi}^2{\longrightarrow}{\infty}$ when
$L{\longrightarrow}{\infty}$ and thus ${\theta}_{eff}$ corresponds
to strong noncommutativity . For $q<\frac{1}{2}$ we find that
${\xi}^2{\longrightarrow}0$ when $L{\longrightarrow}{\infty}$ and
${\theta}_{eff}$ corresponds to weak noncommutativity whereas for
$q=\frac{1}{2}$ the effective noncommutativity parameter is
exactly given by ${\theta}_{eff}^2=2{\theta}^2$ . Let us  point
out here that the above result can also be derived from coherent
states and star products .

We can immediately see from (\ref{action2}) that Lorentz
invariance is here fully maintained at the level of the action in
the form of the explicit rotational $SU(2)$ symmetry of the fuzzy
sphere. The $SO(3)$ symmetry is broken down to $SO(2)$ symmetry
only by the constraint (\ref{lorentz}) . Furthermore the
noncommutativity parameter ${\theta}^2$ from (\ref{cond4}),
(\ref{limit}) and (\ref{lorentz}) provides the only length scale
in the problem and hence $\theta$ for all values of $R$ and $L$
defines the volume and distances of the underlying space-time and
therefore it can not be treated as some dimensionfull coupling
constant in the theory.

\section{The Chern-Simons Action}

It was shown in \cite{alex} that the dynamics  of open strings
moving in a curved space with ${\bf S}^3$ metric in the presence
of a non-vanishing Neveu-schwarz B-field   and with Dp-branes is
not precisely equivalent , to the leading order in the string
tension , to the above gauge theory (\ref{action2}) . This is of
course in contrast with the case of strings in flat backgrounds .
Indeed the effective action turns out to contain also an extra
crucial term given by the Chern-Simons action

\begin{eqnarray}
S_{CS}&=&-\frac{R}{6g_f^2}\frac{1}{L+1}\epsilon_{abc}Tr_fF_{ab}D_c-\frac{1}{6g_f^2}\frac{1}{L+1}Tr_f(D_a^2-\frac{|L|^2}{R^2}).\label{CS}
\end{eqnarray}
From string theory  point of view the most natural candidate for a
gauge action on the fuzzy sphere is therefore given instead by
the action
\begin{eqnarray}
S_{L}&=&S_{L,R}+S_{CS}.\label{cite}
\end{eqnarray}
We remark that the Chern-Simons term vanishes in the planar limit
(\ref{limit}) and thus its addition does not change the argument
of the previous section. This fact can also be seen by rewriting
the Chern-Simons action in terms of the gauge field directly as
follows . We write $D_a=\frac{1}{R}L_a+A_a$ and then compute
\begin{eqnarray}
S_{CS}&=&-\frac{1}{2g_f^2}\frac{1}{L+1}{\epsilon}_{abc}Tr_f\bigg[\frac{1}{2}F_{ab}^{(0)}A_c+\frac{iR}{3}[A_a,A_b]A_c\bigg]=-\frac{1}{2g_f^2}\frac{1}{L+1}Tr_f\bigg[\frac{1}{2}{\epsilon}_{ij3}F_{ij}^{(0)}A_3+R{\epsilon}_{3ij}A_jF_{3i}\bigg]
\end{eqnarray}
where $F_{ab}^{(0)}=i[L_a,A_b]-i[L_b,A_a]+{\epsilon}_{abc}A_c$ ,
$F_{ab}= \frac{1}{R}F_{ab}^{(0)}+i[A_a,A_b]$. Hence in the planar
limit where we can set $A_3=0$ ,
$A_i=\frac{1}{|L|^{q-\frac{1}{2}}}\hat{A}_i$ and
$F_{3i}=\frac{1}{\theta
|L|^{2q-\frac{1}{2}}}{\epsilon}_{ij}\hat{D}_j$ it is quite
obvious that we will have
$S_{CS}=-\frac{1}{2g_f^2}\frac{1}{L+1}Tr_f\frac{\hat{A}_i\hat{D}_i}{|L|^{2q-1}}
$, i.e this action vanishes also as $\frac{1}{L^{2q}}$.

As it turns out however the addition of the Chern-Simons term
simplifies considerably perturbation theory. Indeed one can check
that the quadratic term of the action $S_{L}$ is of the form
\begin{eqnarray}
S_{L}^{(2)}=-\frac{1}{2g_f^2}\frac{1}{L+1}Tr_f\big([L_a,A_b]^2-[L_a,A_a]^2\big).
\end{eqnarray}
In other words and after an obvious gauge fixing the propagator
of the theory is simply given by $\frac{g_f^2}{{\cal L}^2}$ which
is very similar to the propagator on the plane . This
simplification seems to be related to the fact that the action
$S_{L}$ has the extra symmetry
$A_a{\longrightarrow}A_a+{\alpha}_a{\bf 1}_{L+1}$ for any
constants ${\alpha}_a$, in other words it is invariant under
global translations in the space of gauge fields . We choose for
simplicity to fix this symmetry by restricting the gauge field to
be traceless, i.e by removing the zero modes . The action we will
study is therefore given by $S_{L}$ with the constraint
(\ref{constraint1}) and the corresponding partition function is
defined by
\begin{eqnarray}
Z_L[g_f^2{\theta}^2;J]=\int {\prod}_{a=1}^3{\cal D}D_a
~{\delta}\bigg(D_a^2-\frac{|L|^2}{R^2}\bigg)~e^{-S_L-\frac{R}{L+1}Tr_fJ_aD_a}.\label{Z}
\end{eqnarray}
This theory was extensively studied for finite $L$ ( keeping $R$
fixed ) in \cite{denjoydri} . As we have said earlier the
constraint $D_a^2-\frac{|L|^2}{R^2}=0$ simply removes the normal
component of the gauge field  which is defined here by
${\Phi}=\frac{D_a^2-|L|^2}{2|L|}$. In \cite{denjoydri} we have
shown explicitly that  without this constraint the model
(\ref{cite}) has a gauge-invariant UV-IR mixing. Furthermore by
adding a large mass term for the normal component of the gauge
field in the form $M^2Tr_f{\Phi}^2$ we can show that , in the
limit where $M{\longrightarrow}{\infty}$ first ( which will
implement the constraint ) then  $L{\longrightarrow}{\infty}$  ,
the mixing is removed . This result is confirmed by the large $L$
analysis of \cite{steinacker} and suggests that the UV-IR mixing
has its origin in the coupling of extra degrees of freedom to the
theory which are here identified with the scalar normal component
of the gauge field. The other exciting result regarding this
model is the existence of a first order phase transition in the
system at some large coupling between a pure matrix model and a
fuzzy sphere model. This phase transition was  confirmed
numerically by \cite{jun} and suggests that the one-loop quantum
theory is actually an exact result .

\section{$U_{*}(1)$ Theory in $d-$Dimensions} The space ${\bf
R}_{\theta}^d$ in general can be only partially noncommutative ,
i.e the Poisson tensor ${\theta}^2B_{ij}$ is of rank $2r{\leq}d$.
This means in particular that we have only $2r$ noncommuting
coordinates. We will now concentrate on the case of $U(1)$ gauge
theory on a minimal noncommutative space , i.e $r=1$ . The
notation for $i=1,2$ remains $\hat{x}_i$ which correspond in the
star picture to the noncommutative coordinates $x_1$ and $x_2$  (
or equivalently the complex coordinates  $z=x_1+ix_2$ and
$\bar{z}=x_1-ix_2$ ). For $i=3,...,d$ or ${\mu}=1,...,d-2$ we
have the commutative coordinates  $\hat{x}_i{\equiv}x_{\mu}$ .
The commutation relations are therefore

\begin{eqnarray}
[\hat{x}_i,\hat{x}_j]=i{\theta}^2{\epsilon}_{ij}~,~[x_{\mu},x_{\nu}]=[\hat{x}_i,x_{\mu}]=0,\label{140}
\end{eqnarray}
where we have set $B_{12}{\equiv}{\epsilon}_{12}=1$ for
simplicity \cite{ nekrasov} .
The derivatives on this noncommutative space will now be defined
by
\begin{eqnarray}
&&\hat{\partial}_i=-\frac{i}{{\theta}^2}({\epsilon}^{-1})_{ij}\hat{x}_j=-\hat{\partial}_i^{+}~,~[\hat{\partial}_i,\hat{\partial}_j]=\frac{i}{{\theta}^2}({\epsilon}^{-1})_{ij}=-\frac{i}{{\theta}^2}{\epsilon}_{ij}~,~[\hat{\partial}_i,x_{\mu}]=0~,~({\rm
for}~i=1,2)~\nonumber\\
&&\hat{\partial}_i{\equiv}{\partial}_{\mu}=-{\partial}_{\mu}^{+}~,~[{\partial}_{\mu},\hat{x}_i]=0~,~[{\partial}_{\mu},x_{\nu}]={\delta}_{\mu
\nu}~({\rm for}~i=3,...,d~,~\mu=1,...,d-2)~.
\end{eqnarray}
Also we have $[{\partial}_{\mu},\hat{\partial}_i]=0$ , $i=1,2$ .
The covariant derivatives are on the other hand given by
\begin{eqnarray}
&&\hat{D}_i=-i\hat{\partial}_i+\hat{A}_i~,~({\rm
for}~i=1,2)\nonumber\\
&&\hat{D}_i{\equiv}\hat{D}_{\mu}=-i\hat{\partial}_i+\hat{A}_i{\equiv}-i{\partial}_{\mu}+\hat{A}_{\mu}~,~({\rm
for}~i=3,...,d~,~\mu=1,...,d-2).
\end{eqnarray}
Both $\hat{A}_i$ and $\hat{A}_{\mu}$ are still operators , indeed
we can write the Fourier expansion
\begin{eqnarray}
\hat{A}_i{\equiv}\hat{A}_{i}(\hat{x}_1,\hat{x}_2,x_{\mu})=\int
\frac{d^dk}{(2{\pi})^d}\tilde{A}_{i}(k)e^{ik_1\hat{x}_1+ik_2\hat{x}_2}e^{ik_{\mu}{x}_{\mu}}~,~({\rm
for}~{\rm all}~i=1,...,d).
\end{eqnarray}
The operators $\hat{A}_i$'s clearly act on the same Hilbert space
${\bf H}$ on which the coordinate operators $\hat{x}_1$ and
$\hat{x}_2$ act . The operators $\hat{A}_i$'s can be mapped to
the fields $A_i$ given by
\begin{eqnarray}
\hat{A}_i(\hat{x}_1,\hat{x}_2,x_{\mu})=\int d^2x
A_i(x_1,x_2,x_{\mu}){\Delta}(\hat{x}_1,\hat{x}_2,x_1,x_2),
\end{eqnarray}
where the Weyl map is given by
\begin{eqnarray}
{\Delta}(\hat{x}_1,\hat{x}_2,x_1,x_2)=\int
\frac{d^2k}{(2{\pi})^2}e^{ik_i\hat{x}_i}e^{-ik_ix_i}.\label{wm}
\end{eqnarray}
Remark for example that if $\hat{A}_i$ did not depend on the
operators $\hat{x}_1$ and $\hat{x}_2$ then one can simply make
the identification $\hat{A}_i(x_{\mu}){\equiv}A_i(x_{\mu})$ since
$\int d^2x {\Delta}(\hat{x}_1,\hat{x}_2,x_1,x_2)=1$ . Indeed the
star product is given now by
\begin{eqnarray}
f*g(x)=e^{\frac{i}{2}{\theta}^2\sum_{i,j=1}^2{\epsilon}_{ij}\frac{\partial}{{\partial}{\xi}_i}\frac{\partial}{{\partial}{\eta}_j}}f(x+\xi)g(x+\eta)|_{\xi=\eta=0},
\end{eqnarray}
and clearly it involves only the two derivatives
$\frac{\partial}{{\partial}x_1}$ and
$\frac{\partial}{{\partial}x_2}$ so if  both $f$ and $g$ do not
depend on the two coordinates $x_1$ and $x_2$ then
$f*g(x){\equiv}f(x)g(x)$ . In fact even in the case where only
one of the two functions $f$ and $g$ is independent of $x_1$ and
$x_2$ we have $f*g(x){\equiv}f(x)g(x)$ .

The curvature is defined now by
\begin{eqnarray}
&&\hat{F}_{ij}=i[\hat{D}_i,\hat{D}_j]+\frac{1}{{\theta}^2}{\epsilon}_{ij}=[\hat{\partial}_i,\hat{A}_j]-[\hat{\partial}_j,\hat{A}_i]+i[\hat{A}_i,\hat{A}_j]\nonumber\\
&&\hat{F}_{\mu
i}=i[\hat{D}_{\mu},\hat{D}_i]={\partial}_{\mu}\hat{A}_i-[\hat{\partial}_{i},\hat{A}_{\mu}]+i[\hat{A}_{\mu},\hat{A}_i]\nonumber\\
&&\hat{F}_{\mu
\nu}=i[\hat{D}_{\mu},\hat{D}_{\nu}]={\partial}_{\mu}\hat{A}_{\nu}-{\partial}_{\nu}\hat{A}_{\mu}+i[\hat{A}_{\mu},\hat{A}_{\nu}],
\end{eqnarray}
where $i$ above stands for the two values $1$ and $2$ and $\mu$
stands for the rest . Gauge transformations are also operators
$\hat{U}$ which act as usual , namely
\begin{eqnarray}
&&\hat{D}_{\mu}^{U}=\hat{U}\hat{D}_{\mu}\hat{U}^{+}{\longrightarrow}\hat{A}^U_{\mu}=\hat{U}\hat{A}_{\mu}\hat{U}^{+}-i\hat{U}{\partial}_{\mu}(\hat{U}^{+})\nonumber\\
&&\hat{D}_{i}^{U}=\hat{U}\hat{D}_{i}\hat{U}^{+}{\longrightarrow}\hat{A}^U_{i}=\hat{U}\hat{A}_{i}\hat{U}^{+}-i\hat{U}[\hat{\partial}_{i},\hat{U}^{+}]~,\nonumber\
\end{eqnarray}
and hence $\hat{F}_{ij}^U=\hat{U}\hat{F}_{ij}\hat{U}^{+}$ for all
$i,j=1,...,d$ . The Yang-Mills action for $U_{*}(1)$ gauge theory
on ${\bf R}^2_{\theta}{\times}{\bf R}^{d-2}$  is written in this
case as
\begin{eqnarray}
S_{\theta}&=& \frac{{\theta}^2}{4g^2}\int
d^{d-2}x\sum_{i,j=1}^dTr\hat{F}_{ij}^2\nonumber\\
&=&\frac{{\theta}^2}{4g^2}\int d^{d-2}xTr\hat{F}_{\mu
\nu}^2+\frac{{\theta}^2}{2g^2}\int
d^{d-2}x\sum_{i=1}^2Tr\hat{F}_{\mu
i}^2+\frac{{\theta}^2}{4g^2}\int
d^{d-2}x\sum_{i,j=1}^2Tr\hat{F}_{ij}^2.\label{149}
\end{eqnarray}
In above we have deliberately used the fact that we can replace
the integral over the noncommutative directions $x_1$ and $x_2$
by a trace over an infinite dimensional Hilbert space by using the
Weyl Map introduced in (\ref{wm}) . By doing this we have
therefore also replaced the underlying star product of functions
by pointwise multiplication of operators  . The trace $Tr$ in
(\ref{149}) is thus associated with the two noncommutative
coordinates $x_1$ and $x_2$ . It is curious enough however that
the above model looks very much like a $U(\infty)$ gauge theory
on ${\bf R}^{d-2}$ with a Higgs particle in the adjoint of the
group. This is in fact our original motivation for wanting to
regularize the NC plane with a fuzzy sphere .

For each point $x_{\mu}$ of the $(d-2)-$dimensional commutative
submanifold ${\bf R}^{d-2}$ , the action (\ref{149}) is
essentially an infinite dimensional matrix model and hence it can
be regularized and made into a finite dimensional matrix model if
we approximate for example the noncommutative plane by a fuzzy
sphere. As we explained earlier the trace
$\frac{{\theta}^2}{g^2}Tr$ will be replaced by
$\frac{{R}^2}{g_L^2}Tr_f$ where $g_L^2=g_f^2(L+1)$ and the first
two terms in the action  become
\begin{eqnarray}
\frac {{\theta}^2}{4g^2}\int d^{d-2}xTr\hat{F}_{\mu \nu}^2+ \frac{
{\theta}^2}{2g^2}\int d^{d-2}x\sum_{i=1}^2Tr\hat{F}_{\mu
i}^2{\longrightarrow} \frac{1}{4{\lambda}^2}\int d^{d-2}x
Tr_f{\cal F}_{\mu \nu}^2-\frac{1}{2{\lambda}^2}\int
d^{d-2}x\sum_{a=1}^3Tr_f[{\cal D}_{\mu},D_a]^2~,~
\end{eqnarray}
with ${\lambda}^2=\frac{g_L^2}{{R}^2}=\frac{g_f^2(L+1)}{R^2}$ ,
where we have also replaced the operators $\hat{A}_{\mu}=\hat{D}_{\mu}+i{\partial}_{\mu}$ and
$\hat{D}_i$ by the $(L+1){\times}(L+1)$ dimensional matrices
${\cal A}_{\mu}={\cal D}_{\mu}+i{\partial}_{\mu}$ and $D_a$ respectively. In above ${\cal F}_{\mu
\nu}=i[{\cal D}_{\mu},{\cal D}_{\nu}]$ while the index $a$ runs
over $1,2,3$ since the fuzzy sphere is described by a
$3-$dimensional calculus . In here the fuzzy sphere is only
thought of as a regulator of the noncommutative plane which
preserves exact gauge invariance . Classically we have found that
$g_f^2=g^2$ whereas the effective noncommutativity parameter appearing in the Moyal-Weyl action  is  ${\theta}^2{\xi}^2$  . The coupling constant ${\lambda}^2$ has dimension
$M^{6-d}$ whereas the covariant derivatives ${\cal D}_{\mu}$ and
$D_a$ are as before of dimension $M$ . Remark furthermore that in
the continuum planar limit $L,R{\longrightarrow}{\infty}$ keeping
${\theta}$ fixed in which the fuzzy sphere reproduces the
noncommutative plane the combination
${\lambda}^2\frac{|L|^{2q}}{L+1}$ is kept fixed equal to
$\frac{g_f^2}{{\theta}^2}$: this is the analogue of 't Hooft
planar limit in this context.

The last term in (\ref{149}) has the following interpretation .
For each point $x_{\mu}$ of the $(d-2)-$dimensional commutative
submanifold this term is exactly equivalent to a $U(1)$ gauge
theory on a noncommutative ${\bf R}^{2}$ . This term and as we
have explained in previous sections will therefore be regularized
by the sum of the actions (\ref{action2}) +(\ref{CS}) . In terms
of $D_a$ this action reads
\begin{eqnarray}
&&\frac{{\theta}^2}{4g^2}\int
d^{d-2}x\sum_{i,j=1}^2Tr\hat{F}_{ij}^2{\longrightarrow}
-\frac{R^2}{4g_L^2}\int d^{d-2}xV(D_a),
\end{eqnarray}
where
\begin{eqnarray}
&&V(D_a)=Tr_f[D_a,D_b]^2- \frac{4i}{3R}{\epsilon}_{abc}
Tr_f[D_a,D_b]D_c- \frac{2}{3R^4}(L+1)|L|^2.
\end{eqnarray}
As opposed to the case of perturbation theory where the
Chern-Simons term played a crucial role in simplifying the
propagator and as a consequence the model as a whole ,we can see
in here that in the large $R$ limit the Chern-Simons contribution
is rather small compared to the Yang-Mills contribution and hence
this term becomes irrelevant in this limit .

The full regularized action $S_{\theta;L}$ becomes
\begin{eqnarray}
S_{\theta;L}=\frac{1}{4{\lambda}^2}\int d^{d-2}x Tr_f{\cal
F}_{\mu \nu}^2-\frac{1}{2{\lambda}^2}\int
d^{d-2}x\sum_{a=1}^3Tr_f[{\cal
D}_{\mu},D_a]^2-\frac{1}{4{\lambda}^2}\int d^{d-2}xV(D_a).
\end{eqnarray}
The matrices ${\cal A}_{\mu}$ , $D_a$ are of course  still
functions on the commutative part ${\bf R}^{d-2}$ of ${\bf
R}^d_{\theta}$ , i.e ${\cal A}_{\mu}{\equiv}{\cal
A}_{\mu}(x_{\mu})$, $D_a{\equiv}D_a(x_{\mu})$ ,
${\mu}=1,2,...,d-2$. Gauge transformations
$\hat{U}{\equiv}\hat{U}(x_{\mu})$ of the original action are
implemented now by unitary transformations $U{\equiv}U(x_{\mu})$
acting on the $(L+1)-$dimensional Hilbert space of the
irreducible representation $\frac{L}{2}$ of $SU(2)$ which leave
the above action $S_{\theta;L}$ invariant . They are given
explicitly by $D_a{\longrightarrow}{U}D_a{U}^{+}$, ${\cal
D}_{\mu}{\longrightarrow}U{\cal D}_{\mu}U^{+}$ ( or equivalently
${\cal A}_{\mu}{\longrightarrow}{U}{\cal
A}_{\mu}{U}^{+}-iU{\partial}_{\mu}U^{+}$ where ${\cal
D}_{\mu}=-i{\partial}_{\mu}+{\cal A}_{\mu}$). This is clearly a
$U(L+1)$ gauge theory with adjoint matter , i.e the original
noncommutative degrees of freedom are traded for ordinary color
degrees of freedom which in fact resembles very much what happens
on the noncommutative torus under Morita equivalence . In
particular the components of the covariant derivative in the
noncommutative directions, i.e $D_a$ , $a=1,2,3$ , are now simple
scalar fields with respects to the other commutative $d-2$
dimensions. Quantization of the above model with the constraint
(\ref{constraint1}) corresponds therefore to an ordinary quantum
field theory .

\subsection{The Non-Linear Sigma Model}
\subsubsection{Light-Cone Gauge}
We use now the notation $N{\equiv}L+1$ and work in $d=4$ . The
field ${\cal A}_{\mu}$ can be separated into a $U(1)$ gauge
field  $a_{\mu}$ and an $SU(N)$ gauge field $A_{\mu}$ as follows
\begin{eqnarray}
{\cal A}_\mu(x)=a_{\mu}(x){\bf 1}+A_{\mu}(x)~,~A_{\mu}(x)=A_{\mu
A}(x)T_A,
\end{eqnarray}
where we have introduced the $SU(N)-$Gell-Mann matrices
$T_A=\frac{{\lambda}_A}{2}$ , $A=1,...,N^2-1$ , which satisfy the
usual conditions
\begin{eqnarray}
T_A^{+}=T_A~,~Tr_fT_A=0~,~Tr_fT_AT_B=\frac{1}{2}{\delta}_{AB}~,~[T_A,T_B]=if_{ABC}T_C.
\end{eqnarray}
The curvature becomes
\begin{eqnarray}
&&{\cal F}_{\mu \nu}=f_{\mu \nu }+F_{\mu \nu}\nonumber\\
&&f_{\mu \nu }={\partial}_{\mu}a_{\nu }-{\partial}_{\nu}a_{\mu
}\nonumber\\
&&F_{\mu
\nu}={\partial}_{\mu}A_{\nu}-{\partial}_{\nu}A_{\mu}+i[A_{\mu},A_{\nu}]=F_{\mu
\nu C}T_C~,~F_{\mu \nu C}={\partial}_{\mu}A_{\nu
C}-{\partial}_{\nu}A_{\mu C}-f_{ABC}A_{\mu A}A_{\nu B}.
\end{eqnarray}
The first term in the action becomes
\begin{eqnarray}
\frac{1}{4{\lambda}^2}\int d^{d-2}xTr_f{\cal F}_{\mu
\nu}^2=\frac{1}{4{\lambda}^2}\int d^{d-2}x\bigg[ Nf_{\mu \nu
}^2+Tr_fF_{\mu \nu }^2\bigg].\label{pure}
\end{eqnarray}
Similarly the gauge transformation ${U}=exp(i{\Lambda}(x))$ ,
where ${\Lambda}$ is a general $N{\times}N$ matrix , splits into
a $U(1)$ gauge transformation and an $SU(N)$ gauge
transformation, i.e
${U}{\equiv}e^{i{{\Lambda}(x)}}=e^{i({\alpha}(x)+{\beta}(x))}={W}(x).V(x)$
where $W(x)=e^{i{\alpha}(x)}$ and $V(x)=e^{i{\beta}(x)}$ with
$\alpha(x)$ a function on ${\bf R}^{d-2}$ and
${\beta}(x)={\beta}_A(x)T_A$  . Hence the gauge transformation
${\cal A}_{\mu}{\longrightarrow}{\cal A}_{\mu}^{U}={U}{\cal
A}_{\mu}{U}^{+}-i{U}{\partial}_{\mu}({U}^+)$ takes now the form
\begin{eqnarray}
&&a_{\mu }{\longrightarrow}a_{\mu }^U=a_{\mu
}-i{W}({\partial}_{\mu}{W}^{+})~,~A_{\mu}{\longrightarrow}A_{\mu}^U=VA_{\mu}V^{+}-iV({\partial}_{\mu}V^{+}).
\end{eqnarray}
Similarly we write
\begin{eqnarray}
D_a={n}_{a}+{\Phi}_a~,~{\Phi}_a={\Phi}_{aA}T_A,
\end{eqnarray}
In this case the gauge transformation
$D_a{\longrightarrow}D_a^{U}={U}D_a{U}^{+}$ becomes
\begin{eqnarray}
&&{n}_{a}{\longrightarrow}{n}_{a}^U={n}_{a}~,~{\Phi}_{a}{\longrightarrow}{\Phi}_{a}^U=V{\Phi}_aV^{+}.
\end{eqnarray}
This means that ${\Phi}_a$ is a scalar field (``scalar'' with
respect to the commutative directions of ${\bf R}^d_{\theta}$ )
which transforms in the adjoint representation of the non-abelian
subgroup $SU(N)$ of $U(N)$. In fact  $n_a$ is also a scalar field
in the same sense . The action takes now the explicit form
\begin{eqnarray}
S_{\theta;L}&=&\frac{1}{4{\lambda}^2}\int d^{d-2}x Tr_f{F}_{\mu
\nu}^2+\frac{N}{4{\lambda}^2}\int d^{d-2}f^2_{\mu
\nu}-\frac{1}{2{\lambda}^2}\int
d^{d-2}xTr_f[{D}_{\mu},{\Phi}_a]^2+\frac{N}{2{\lambda}^2}\int
d^{d-2}x({\partial}_{\mu}{n}_a)^2\nonumber\\
&-&\frac{1}{4{\lambda}^2}\int d^{d-2}xV({\Phi}_a),\label{163}
\end{eqnarray}
where $D_{\mu}=-i{\partial}_{\mu}+A_{\mu}$ and
$V({\Phi}_a)=Tr_f[{\Phi}_a,{\Phi}_b]^2-\frac{4i}{3R}{\epsilon}_{abc}Tr_f[{\Phi}_a,{\Phi}_b]{\Phi}_c-\frac{2}{3{\theta}^4}\frac{L+1}{|L|^2}$
. The second term in this  action is trivial decribing an abelian
$U(1)$ gauge field on an Euclidean $(d-2)-$dimensional flat
spacetime with no interactions with the other fields. In the case
of $d=4$ the non-abelian part of the action ( i.e the first term
in (\ref{163}) ) is seen to be defined on a two dimensional
spacetime and thus it can be simplified further if one uses the
light-cone gauge \cite{thooft}  . To this end we rotate first to
Minkowski  signature then we fix the $SU(L+1)$ symmetry by going
to the light-cone gauge given by $A_1=A_2=\sqrt{2}{\lambda}A_{+}$
( this is equivalent to $A_{-}=0$ ) . Similarly we  fix the $U(1)$
gauge symmetry by writing $ a_{\mu
}=\sqrt{\frac{2}{N}}\lambda({\epsilon}_{\mu
\lambda}{\partial}^{\lambda}{\sigma}+{\partial}_{\mu}{\eta})$ .
The action becomes therefore
\begin{eqnarray}
exp(iS_{\theta ;L})&=&expi\bigg(-\int d^{2}x
({\partial}^2{\sigma})^2-\int
d^{2}x({\partial}_{-}A_{+A})^2+\frac{N}{2{\lambda}^2}\int
d^{2}x({\partial}_{\mu}n_a)({\partial}^{\mu}n_a)+\frac{1}{4{\lambda}^2}\int
d^{2}x({\partial}_{\mu}{\Phi}_{aA})({\partial}^{\mu}{\Phi}_{aA})\nonumber\\
&-&\frac{1}{\lambda}f_{ABC}\int
d^2x({\partial}_{-}{\Phi}_{aA})A_{+B}{\Phi}_{aC}-\frac{1}{4{\lambda}^2}\int
d^{2}xV({\Phi}_a)\bigg).
\end{eqnarray}
Remark from above that there is no ghost term in the light-cone
gauge \cite{thooft}. The partition function is of the form
\begin{eqnarray}
Z&=&\int {\cal D}{\sigma}{\cal D}A_{+A}{\cal D}n_a{\cal
D}{\Phi}_{aA}e^{iS_{\theta;L}}{\delta}(D_a^2-\frac{|L|^2}{R^2}).
\end{eqnarray}
The delta function is clearly inserted in order to implement the
constraint (\ref{constraint1}). It is rather trivial to see that
the field $\sigma$ is completely decoupled from the rest of the
dynamics and so it simply drops out from the action whereas we
notice that we can perform the integral over the $A_{+}$ fields
in a straightforward manner to give a non-local Coulomb
interaction between the ${\Phi}_{aC}$ fields . We define $
f_{ABC}({\partial}_{-}{\Phi}_{aA}){\Phi}_{aC}{\equiv}(\vec{\Phi}_a{\times}_L{\partial}_{-}\vec{\Phi}_a)_B$
and then write the final result  in the form
\begin{eqnarray}
\hat{S}_{\theta;L}&=&\frac{N}{2{\lambda}^2}\int
d^{2}x({\partial}_{\mu}n_a)({\partial}^{\mu}n_a)+\frac{1}{4{\lambda}^2}\int
d^{2}x({\partial}_{\mu}{\Phi}_{aA})({\partial}^{\mu}{\Phi}_{aA})\nonumber\\
&-&\frac{1}{4{\lambda}^2}\int d^2xd^2y
(\vec{\Phi}_a{\times}_L{\partial}_{-}\vec{\Phi}_{a})_A(x)D_{AB}^{-1}(x,y)(\vec{\Phi}_b{\times}_L{\partial}_{-}\vec{\Phi}_{b})_B(y)-\frac{1}{4{\lambda}^2}\int
d^{2}xV({\Phi}_a).\label{174}
\end{eqnarray}
$D_{AB}^{-1}(x,y)$ is the propgator of the $A_{+A}$ fields , i.e $D^{-1}_{AB}(x,y)=-\frac{{\delta}_{AB}}{2{\pi}}|x_{-}-y_{-}|{\delta}(x_{+}-y_{+})$.

\subsubsection{The Constraint}

Next we analyze the constraint $D_aD_a=\frac{|L|^2}{R^2}$ . This
can be rewritten in the form
\begin{eqnarray}
&&{n}^2_{a}+\frac{1}{2N}{\Phi}_{aA}^2=\frac{|L|^2}{{R}^2}~,~{n}_{a}{\Phi}_{aC}+\frac{1}{4}d_{ABC}{\Phi}_{aA}{\Phi}_{aB}=0,
\end{eqnarray}
where we have used the identities
$T_AT_B=\frac{1}{2N}{\delta}_{AB}+\frac{1}{2}(d_{ABC}+if_{ABC})T_C$
, $ TrT_AT_BT_C=\frac{1}{4}(d_{ABC}+if_{ABC})$. From the
structure of this constraint and from the action (\ref{174}) we
can see that the field $n_a$ appears at most quadratically and
hence it can be integrated out without much effort . The relevant
part of the partition function reads
\begin{eqnarray}
Z_{\vec{n}}&=&\int {\cal
D}n_a~exp\bigg(-\frac{N}{2{\lambda}^2}\int d^2x
({\partial}_{\mu}n_a)^2\bigg)~{\delta}\big(n_a^2+\frac{1}{2N}{\Phi}_{aA}^2-\frac{1}{{\theta}^2}\big)~{\delta}\big(n_a{\Phi}_{aC}+\frac{1}{4}d_{ABC}{\Phi}_{aA}{\Phi}_{aB}\big)\nonumber\\
&=&\int {\cal D}n_a~ {\cal D}J~ {\cal
D}J_C~exp\bigg(-\frac{N}{2{\lambda}^2}\int d^2x
({\partial}_{\mu}n_a)^2+iJ(n_a^2+\frac{1}{2N}{\Phi}_{aA}^2-\frac{|L|^2}{R^2})+iJ_C(n_a{\Phi}_{aC}+\frac{1}{4}d_{ABC}{\Phi}_{aA}{\Phi}_{aB})\bigg).\nonumber\\
\end{eqnarray}
The delta functions which are obviously enforcing the constraint
are represented for convenience with Lagrange multiplier fields
$J$ and $J_C$ . In the above partition function $Z_{\vec{n}}$ we
have also rotated back to Euclidean signature for ease of
manipulations . The equations of motion read as follows
\begin{eqnarray}
{\partial}^2n_a=-\frac{2i{\lambda}^2}{N}(Jn_a+\frac{1}{2}J_C{\Phi}_{aC}).
\end{eqnarray}
Writing now $n_a=e_a+q_a$ where the fixed background field $e_a$
is assumed to solve the above equations of motion whereas $q_a$
is the fluctuation field one can compute in a straightforward
manner the partition function
\begin{eqnarray}
Z_{\vec{n}}={\delta}\bigg(\frac{1}{2}e_a{\Phi}_{aC}+\frac{1}{4}d_{ABC}{\Phi}_{aA}{\Phi}_{aB}\bigg)~\int
{\cal D}J ~exp\bigg(\frac{3}{2}\int d^2x
<x|Log\big[{\partial}^2+\frac{2i{\lambda}^2}{N}J\big]|x>+i\int
d^2xJ(\frac{1}{2N}{\Phi}_{aA}^2-\frac{|L|^2}{{R}^2})\bigg).\nonumber\\
\end{eqnarray}
In the large $L$ limit the exact quantum result $\frac{3}{2}\int
d^2x <x|Log\big[{\partial}^2+\frac{2i{\lambda}^2}{N}J\big]|x>$
becomes independent of $J$ and hence the above partition function
reduces simply to a product of two delta functions , namely $
Z_{\vec{n}}={\delta}\big(\frac{1}{2}e_a{\Phi}_{aC}+\frac{1}{4}d_{ABC}{\Phi}_{aA}{\Phi}_{aB}\big){\delta}\big(\frac{1}{2N}{\Phi}_{aA}^2-\frac{|L|^2}{{R}^2}\big)$
where now $e_a$ is the solution of the equation
${\partial}^2e_a{\longrightarrow}0$ . In other words the
integration over the field $n_a$ in the large $L$ limit is
essentially equivalent to imposing on the field
${\chi}_{aA}=\frac{R}{|L|}\frac{1}{\sqrt{2N}}{\Phi}_{aA}$ the
following constraint
\begin{eqnarray}
&&{\chi}_{aA}^2=1~,~d_{ABC}{\chi}_{aA}{\chi}_{aB}=-\frac{2e_aR}{|L|\sqrt{2(L+1)}}{\chi}_{aC}.\label{con}
\end{eqnarray}
From the above derivation this result clearly does not depend on
the metric we used and so it must also be valid for Minkowski
signature . Since in the limit the vector $e_a$ is an arbitrary
solution of ${\partial}^2e_a=0$ we take it for simplicity
$x-$independent . The reduced action becomes on the other hand
\begin{eqnarray}
\bar{S}_{\theta;L}&=&\frac{1}{4{\bar{\lambda}}^2}\int
d^{2}x({\partial}_{\mu}{\chi}_{aA})({\partial}^{\mu}{\chi}_{aA})-\frac{|L|^2(L+1)}{2{\bar{\lambda}}^2{R}^2}\int
d^2x \bar{V}({\chi}_a),\label{finalaction}
\end{eqnarray}
where
\begin{eqnarray}
\bar{V}({\chi}_a)&=&\int d^2y
(\vec{\chi}_a{\times}_L{\partial}_{-}\vec{\chi}_{a})_A(x)D_{AB}^{-1}(x,y)(\vec{\chi}_b{\times}_L{\partial}_{-}\vec{\chi}_{b})_B(y)+Tr_f[{\chi}_a,{\chi}_b]^2-\frac{4i}{3|L|\sqrt{2(L+1)}}{\epsilon}_{abc}Tr_f[{\chi}_a,{\chi}_b]{\chi}_c\nonumber\\
&-&\frac{1}{6|L|^2(L+1)}
.
\end{eqnarray}
In here ${\bar{\lambda}}^2=\frac{g_f^2}{2|L|^2}$ . Since
$R^2={\theta}^2|L|^{2q}$ the coupling in front of the potential
$\bar{V}$ behaves in the limit as
$\frac{|L|^2(L+1)}{2\bar{\lambda}^2R^2}{\sim}\frac{1}{{\bar{\lambda}}^2{\theta}^2}(\frac{L}{2})^{3-2q}$
and thus for all scalings with $q>\frac{3}{2}$ this potential
term can be neglected compared to the kinetic term and one ends
up with the following partition function
\begin{eqnarray}
Z=\int {\cal
D}{\chi}_{aA}{\delta}({\chi}_{aA}^2-1){\delta}\bigg(d_{ABC}{\chi}_{aA}{\chi}_{aB}+2{\theta}|L|^{q-\frac{3}{2}}e_a{\chi}_{aA}\bigg)e^{-\bar{S}_{\theta;L}}.
\end{eqnarray}
As we have discussed earlier the fuzzy theory for these particular
scalings becomes a theory living on a noncommutative plane with
effective deformation parameter given by
${\theta}_{eff}^2{\sim}2{\theta}^2(\frac{L}{2})^{2q-1}$ ( see
equation (\ref{effe})) . We are therefore probing the strong
noncommutativity region of the Moyal-Weyl model . The above
partition function can be easily computed   and one finds
\begin{eqnarray}
Z=\int {\cal D}J{\cal D}J_C e^{i \int d^2x J }
exp\bigg({-\frac{3}{2}TRlogD }
\bigg)exp\bigg(-{\vec{e}^2{\theta}^2|L|^{2(q-\frac{3}{2})}\int
d^2x d^2y J_A(x)D^{-1}_{AB}(x,y)J_B(y)}\bigg),
\end{eqnarray}
where $D(=D_{AB}(x,y))$ is now the Laplacian
\begin{eqnarray}
D_{AB}(x,y)={\delta}^2(x-y)\bigg(-\frac{1}{4{\bar{\lambda}}^2}{\partial}^2{\delta}_{AB}+iJ{\delta}_{AB}+iJ_Cd_{ABC}\bigg).
\end{eqnarray}
At this stage it is obvious that in the large $L$ limit only
configurations where $J_A=0$ are relevant and thus one ends up
with the partition function

\begin{eqnarray}
Z=\int {\cal D}J e^{i \int d^2x
J-\frac{3}{2}TRlogD}~,~D_{AB}(x,y)={\delta}^2(x-y)\bigg(-\frac{1}{4{\bar{\lambda}}^2}{\partial}^2{\delta}_{AB}+iJ{\delta}_{AB}\bigg)
\end{eqnarray}
This is exactly the partition function of an $O(M)$ non-linear
sigma model in the limit $M{\longrightarrow}{\infty}$ with
$\bar{\lambda}^2M$ held fixed equal to $6g_f^2$ where
$M=3(N^2-1)=12|L|^2$ . Indeed we have
\begin{eqnarray}
Z=\int {\cal D}J exp\bigg({\frac{i}{4\bar{\lambda}^2} \int d^2x
J-\frac{M}{2}\int d^2x <x|log\big(-{\partial}^2+iJ\big)|x>}\bigg).
\end{eqnarray}
All terms in the exponent are now of the same order $M$ and thus
the model can  be solved using the method of steepest descents .
Minimizing the exponent yields the equation
\begin{eqnarray}
<x|\frac{1}{-{\partial}^2+iJ}|x>=\frac{1}{12g^2_f}.
\end{eqnarray}
Solutions $J(x)$ of this equation are obviously given by
$J(x)=-im^2$ where $m^2$ are positive real constant numbers  and
thus this equation , which reads ( in momentum space ) $\int
\frac{d^2k}{(2{\pi})^2}\frac{1}{k^2+m^2}=\frac{1}{12g_f^2}$ ,
admits the solutions
\begin{eqnarray}
\frac{1}{12g_f^2}=\frac{1}{2{\pi}}log\frac{\Lambda}{m}
\end{eqnarray}
 where we
have also regulated the integral with a momentum cutoff
${\Lambda}>>m$ . In order to get a regulator-independent coupling
constant we will need to renormalize this theory and thus
introduce explicitly a renormalization scale $\mu$   . This is
achieved by the simple prescription
\begin{eqnarray}
\frac{1}{12g_f^2}=\frac{1}{12g_{r}^2}+\frac{1}{2{\pi}}log\frac{\Lambda}{\mu}.
\end{eqnarray}
From this result we can derive the beta function of the theory which we find
to be given by
\begin{eqnarray}
{\beta}(g_r)=\mu\frac{{\partial}g_r}{{\partial}\mu}=-\frac{3}{\pi}g_r^3.\label{beta1}
\end{eqnarray}
This result  , up to a numerical factor ( which can always be
understood as a normalization of the coupling constant ) , is the
same as the result (\ref{beta}) obtained in ordinary one-loop
perturbation theory of the original Moyal-Weyl Plane . The crucial
difference in this case is the fact that the above result is
actually exact to all orders in $\bar{\lambda}^2M=6g_f^2$ and
thus it is intrinsically nonperturbative  \cite{sigma} . The
arbitrariness of the definition of the renormalized coupling
constant is reflected in the fact that the solution of this
theory depends on an arbitrary renormalization mass scale $\mu$ .
Indeed it is not difficult to find that $m=m(g_r,\mu)$ is given by
\begin{eqnarray}
m(g_r^2,\mu)=\mu e^{-\frac{\pi}{6g_r^2}}.\label{mass}
\end{eqnarray}
It is worth pointing out that this mass satisfies the
Callan-Symanzik equation $\big[\mu\frac{\partial}{\partial
\mu}+{\beta}(g_r)\frac{\partial}{\partial g_r}\big]m(g_r^2,\mu)=0$
and hence everything is under control. Finally the non-vanishing
of this $O(M)-$mass means in particular that this $O(M)$ symmetry
is never broken for all values of $g_r^2$.

\section{Conclusion}

As we have discussed in this paper there are few  problems with
the path integral of field theory on  the canonical
noncommutative Moyal-Weyl spaces . The noncommutative plane is
actually a zero-dimensional matrix model and not a continuum
space. It acts however on an infinite dimensional Hilbert space
and thus we are integrating in the path integral over infinite
dimensional matrices which is a rather formal procedure . The
second problem is the absence of rotational invariance due to the
non-zero value of theta ; the noncommutativity parameter . A
third problem is the appearance in the theory  of a dimensionfull
parameter , this same ${\theta}$ , which goes against the
intuitive argument for this theory to be renormalizable .

The fuzzy sphere is a $0$-dimensional
 matrix model with a
gauge-invariant , Lorentz-invariant UV as well as IR cutoffs . In
this approximation the noncommutative Moyal-Weyl planes can be
simply viewed as large spheres ( i.e with large radii $R$ ) which
are represented by large but finite matrices (i.e with large
representations $L$ of $SU(2)$) . The relevant limit is a double
scaling continuum planar limit where
 for example  the ratio $R/L$ is kept fixed equal to ${\theta}/2$
which is to be identified with the noncommutativity parameter . In
this formulation it is obvious that the noncommutativity
parameter ${\theta}^2$ acquires its dimension of $(\rm length)^2$
from the large radius of the underlying fuzzy approximation and
hence renormalizability is not necessarily threatened .

In this article the above prescription is applied  to
$4-$dimensional noncommutative $U(1)$ gauge theory with some
remarkable results . For simplicity we have considered a minimal
noncommutative space ${\bf R}^2{\times}{\bf R}^2_{\theta}$ . If we
approximate this noncommutative spatial slice ${\bf
R}^2_{\theta}$ by a fuzzy sphere of matrix size $L$ and radius
$R$ as explained above then the noncommutative degrees of freedom
are converted into color degrees of freedom  . Classically it is
seen that the field theory on the fuzzy space ${\bf
R}^2{\times}{\bf S}^2_L$ reduces to the field theory on the
Moyal-Weyl plane ${\bf R}^2{\times}{\bf R}^2_{\theta}$ in the
flattening continuum planar limits $R,L{\longrightarrow}{\infty}$
where $R^2/(|L|^2)^q={\theta}^2$ . The effective noncommutativity
parameter is however  found to be given by
${\theta}_{eff}^2{\sim}2{\theta}^2(\frac{L}{2})^{2q-1}$. In the
quantum theory it turns out that this prescription is also
equivalent to a dimensional reduction of the model where the
noncommutative $U(1)$ gauge theory  in $4$ dimensions is shown to
be equivalent in the large $L$ limit to an ordinary $O(M)$
non-linear sigma model in $2$ dimensions where $M=12|L|^2$ . More
precisely the large $L$ flattening planar limit is proven to be
the same as t'Hoodt limit of the $O(M)$ sigma model in which the
coupling constant $\bar{\lambda}{\longrightarrow}0$ such that
$M\bar{\lambda}^2$ is kept fixed equal to $6g_f^2$ where $g_f$ is
precisely the coupling constant of the original $U(1)$ theory .
This result is only true for the class of scalings in which
$q>\frac{3}{2}$ and where the corresponding Moyal-Weyl plane is
strongly noncommuting  . The model defined this way is also seen
to be an ordinary renormalizable theory which can be solved
exactly using the method of steepest descents to yield the beta
function (\ref{beta1}) . This beta function (\ref{beta1}) agrees
with the one-loop perturbative result (\ref{beta}) but as we have
shown it is also an exact representation of the beta function of
the theory to all orders in $g_r^2$ .

As we have said above the model can be solved exactly in the
large $L$ limit and one finds for a fixed  renormalization scale
$\mu$ and  a fixed renormalized coupling $g_r$ ( or equivalently a
fixed cut-off ${\Lambda}$ and a fixed bare coupling $g_f$ )  a
non-zero $O(M)-$symmetric mass for the different $M$ components
of the sigma model field given by equation (\ref{mass}). This is
clearly non-zero for all values of $g_r^2$ and hence the $O(M)$
symmetry is never broken  in this solution .

Finally from the action (\ref{finalaction}) and from equation
(\ref{effe}) we conclude that for the scalings
$\frac{1}{2}<q<\frac{3}{2}$ we have a strongly noncommuting
Moyal-Weyl plane where the action is dominated by the potential
term , i.e the quantum description in this case is  purely in
terms of a matrix model . For $q<\frac{1}{2}$ the action is still
dominated by the potential term but the Moyal-Weyl plane is
weakly noncommuting . The values $q=\frac{1}{2}$ and
$q=\frac{3}{2}$ are special. For $q=\frac{1}{2}$ the
noncommutativity parameter is given by
${\theta}_{eff}^2=2{\theta}^2$ and the action is dominated by the
potential term whereas for $q=\frac{3}{2}$ the Moyal-Weyl plane
is strongly noncommuting but now both terms in the action
(\ref{finalaction}) are important . The precise meaning of all
this is still not clear .

Including non-trivial field configurations , such as those
introduced in \cite {1} , is still however an open question .
Fermions and as a consequence chiral symmetry , in the sense of
\cite{2,local} , are also not obvious how to formulate in this
limit . Also since the fuzzy sphere parameter $L$ is meant to be a
cut-off we can ask the question how does the theory actually
depends on $L$ , in particular renormalizability of the
$L=\infty$ is an open question. This is obviously a much harder
question and we are currently contemplating adapting the
Polchinski approach to this problem . In $4-$dimensions other
choices for the fuzzy underlying manifolds are available such as
fuzzy  ${\bf CP}^2$ and fuzzy ${\bf S}^4$ but fuzzy ${\bf
S}^2{\times}{\bf S}^2$ seems much more practical as all the
computation in the corresponding QFT's only involve the well
known $SU(2)$ Clebsch-Gordan coefficients \cite{5,7}.

\bibliographystyle{unsrt}

\end{document}